\documentclass[aps,prd,reprint,preprintnumbers]{revtex4-1}
\pdfoutput=1
\usepackage{url}
\usepackage{textcase}
\usepackage{bm}
\usepackage{amsmath}
\usepackage{amsfonts}
\usepackage{amssymb}
\usepackage{hyperref}
\usepackage[table]{xcolor}
\usepackage{multirow}
\usepackage{bbm}
\usepackage{graphicx}
\usepackage{tensor}
\usepackage[caption=false]{subfig}
\usepackage{array}
\usepackage[normalem]{ulem}
\usepackage{float}
\usepackage{fancyhdr}
\usepackage[caption=false]{subfig}

\usepackage{pdfpages}
\makeatletter
\AtBeginDocument{\let\LS@rot\@undefined}
\makeatother

\usepackage{natbib}
\let\Re\relax \DeclareMathOperator{\Re}{Re}
\let\Im\relax \DeclareMathOperator{\Im}{Im}
\renewcommand{\vec}[1]{\boldsymbol{#1}}

\begin{document}

\title{On the landscape of scale invariance in quantum mechanics}

\date{\today}
\author{Daniel K.~Brattan}
\email{danny.brattan@gmail.com}
\affiliation{Interdisciplinary Center for Theoretical Study, University of Science and Technology of China, 96 Jinzhai Road, Hefei, Anhui, 230026 PRC.}
\author{Omrie Ovdat}
\email{somrie@campus.technion.ac.il}
\affiliation{Department of Physics, Technion Israel Institute of Technology, Haifa 3200003, Israel.}
\author{Eric Akkermans}
\email{eric@physics.technion.ac.il}
\affiliation{Department of Physics, Technion Israel Institute of Technology, Haifa 3200003, Israel.}

\begin{abstract}
{\ We consider the most general scale invariant radial Hamiltonian allowing for anisotropic scaling between space and time. We formulate a renormalisation group analysis of this system and demonstrate the existence of a universal quantum phase transition from a continuous scale invariant phase to a discrete scale invariant phase. Close to the critical point, the discrete scale invariant phase is characterised by an isolated, closed, attracting trajectory in renomalisation group space (a limit cycle). Moving in appropriate directions in the parameter space of couplings this picture is altered to one controlled by a quasi periodic attracting trajectory (a limit torus) or fixed points. We identify a direct relation between the critical point, the renormalisation group picture and the power laws characterising the zero energy wave functions.}
\end{abstract}

\preprint{USTC-ICTS-18-08} 

\maketitle

% Why renormalisation group flow is important; in particular fixed pts and why scale invariance matters for fixed pts.
{\ Classical symmetries broken at the quantum level are termed anomalous. Since their discovery \cite{PhysRev.177.2426,Bell:1969ts,PhysRevD.34.674,1993AmJPh..61..142H}, anomalies have become a very active field of research in physics. One class of anomalies describes the breaking of continuous scale invariance (CSI). In the generic case, quantisation of a classically scale invariant Hamiltonian is ill-defined and necessitates the introduction of a regularisation scale \cite{COLEMAN1971552} which breaks CSI altogether. Recently, a sub-class of scale anomalies has been discovered in which a residual discrete scale invariance (DSI) remains after regularisation. Models exhibiting this phenomenon include a non-relativistic particle in the presence of an attractive, inverse square radial potential $\hat{H}_S= p^2/2m - \lambda/r^2$ \cite{Case1950,deAlfaro1976,landau1991quantum,Camblong:2000ec,PhysRevD.68.025006,BraatenPhysRevA.70.052111,HammerSwingle2006a,Kaplan:2009kr,Moroz:2009nm}, the charged and massless Dirac fermion in an attractive Coulomb potential $\hat{H}_D = \gamma^0 \gamma^j p_j - \lambda/r$ \cite{ovdat2017observing} and a class of one dimensional Lifshitz scalars \cite{Alexandre:2011kr} with $\hat{H}_L = \left(p^2/2m \right)^N - \lambda/x^{2N}$ \cite{Brattan:2017yzx}. Any system described by these classically scale invariant Hamiltonians exhibits an abrupt transition in the spectrum at some $\lambda = \lambda_c$. For $\lambda < \lambda_c$, the spectrum contains no bound states close to $E=0$, however, as $\lambda$ goes above $\lambda_c$, an infinite sequence of bound (quasi bound for $\hat{H}_D$) states appears. In addition, in this ``over-critical'' regime, the states surprisingly form a geometric sequence 
  \begin{equation}
    \label{Eq:geometric spectrum}
    E_n = E_0 \exp{(-n \alpha/\sqrt{\lambda-\lambda_{c}})},
  \end{equation}
accumulating at $E = 0$ where $n \in \mathbb{Z}$, $\alpha >0$ and $E_0$ is a number that depends on the regularisation. The existence and structure of the levels is `universal', that is, it does not rely on the details of the potential close to its source. This feature is a signature of residual DSI since $\{E_n\} \rightarrow \{\exp{(-2\pi/\sqrt{\lambda-\lambda_{c}}))} E_n\} = \{E_{n}\}$. Thus, a quantum phase transition occurs at $\lambda_c$ between a continuous scale invariant (CSI) phase and a discrete scale invariant phase (DSI). This transition has been associated with Berezinskii-Kosterlitz-Thouless (BKT) transitions \cite{PhysRevB.46.12664,Kaplan:2009kr,Jensen:2010ga,Jensen:2010vx,Jensen:2011af,2014JSP...156..268D,Gies:2015hia} and has found applications in the Efimov effect \cite{efimov1970energy,Efimov1971,Braaten2006259}, graphene \cite{ovdat2017observing}, QED3 \cite{HerbutPhysRevD.94.025036} and other phenomena \cite{levy1967electron,PhysRevB.46.12664,PhysRevD.48.5940,CamblongEpeleFanchiottiEtAl2001,nisoli2014attractive,Govindarajan:2000ag,Camblong:2003mz,Bellucci200399,Brattan:2018sgc}.}

{\ A useful tool in the characterisation of this phenomenon is the renormalisation group (RG) \cite{WILSON197475}. For the case of $\hat{H}_{S,D,L}$, it consists of introducing an initial short distance scale $L$ and defining model dependent parameters such as $\lambda$, and the boundary conditions, according to physical information. At low energies with respect to the cut-off $L$, a RG formalism allows one to determine the dependence of these parameters on $L$ and thus how physical, regularisation independent, information can be extracted from a scheme dependent result.  For example, an attractive fixed point represents a class of parameters describing the same low energy predictions, characterised by the effective Hamiltonian corresponding to the fixed point. In that sense, the fixed point Hamiltonian describes universal physics. However, termination at a fixed point is not the only possible outcome of a RG flow. In principle, there are three other distinct behaviours that one can find: limit cycles, limit tori and strange attractors \cite{cambel1993applied}; all of which are rare in applications of RG.}

\begin{table*}[!t]
  \centering
  \caption{Summary of the relation between the distinct RG flows describing Hamiltonians \eqref{Eq:GenericHamiltonian0} and the power laws characterising the zero energy wave functions.}
  \smallskip
  \begin{ruledtabular}
  \begin{tabular}{p{4.5cm}p{5cm}p{7cm}}
    \multicolumn{2}{c}{Conditions on $\left\{ \Delta_{i} \right\}$} & Characteristic RG picture \\ \hline
    all roots on symmetry line ($\Re \left[ z \right] = N-1/2$) & $ \Im \left[\Delta_{i}\right]/\Im \left[\Delta_{j}\right] \in \mathbbm{Q}$ & RG space filled by many limit cycles (fig.~\ref{Fig:limitcycle2}) with no fixed points \\
 			      & $ \Im \left[\Delta_{i}\right]/\Im \left[\Delta_{j}\right] \notin \mathbbm{Q}$ & RG space filled by many limit tori with no fixed points \\ \hline
	  some roots off the symmetry line ($\Re \left[ z \right] = N-1/2$) & for roots with $\Re \left[ \Delta_{i} \right] = N-1/2$ if \phantom{space} $ \Im \left[\Delta_{i}\right]/\Im \left[\Delta_{j}\right] \in \mathbbm{Q}$ & isolated limit cycles (fig.~\ref{Fig:limitcycle}) with no fixed points \\
	 			       & for roots with $\Re \left[ \Delta_{i} \right] = N-1/2$ if \phantom{space} $ \Im \left[\Delta_{i}\right]/\Im \left[\Delta_{j}\right] \notin \mathbbm{Q}$ & isolated limit torus (fig.~\ref{Fig:torus}) with no fixed points \\ \hline
    no roots on symmetry line ($\Re \left[ z \right] = N-1/2$) & ~ & $2^N$ fixed points
  \end{tabular}
  \end{ruledtabular}
  \label{tab:features}
  \vspace{-1em}
\end{table*}

{\ The study of $\hat{H}_S$ and $\hat{H}_D$ using RG \cite{BraatenPhysRevA.70.052111,PhysRevB.46.12664,mueller2004renormalization,Kaplan:2009kr,GorskyPhysRevD.89.061702,NishidaPhysRevB.94.085430} shows that the quantum critical phase transition is characterised by two fixed points (UV and IR) for $\lambda < \lambda_{c}$ which combine and annihilate at $\lambda = \lambda_{c}$. For $\lambda > \lambda_{c}$ all the flows are log-periodic in the cut-off and therefore exhibit DSI, independent of the choice of initial boundary condition and scale. The meaning is that for every choice of initial $L$  and boundary condition, there is an infinite equivalent set of scales described by a geometric ladder. This is manifested in \eqref{Eq:geometric spectrum} as it implies $E_{n+k+1}/E_{n+k} = E_{n+1}/E_{n}$ for all $n,k \in \mathbb{Z}$. Remarkably, even in the absence of fixed points, there is universal information in this regime represented by the geometric series factor $E_{n+1}/E_n$.} 

{\ Hamiltonians $\hat{H}_{S,D,L}$ share the property of scaling uniformly under $r \mapsto \Lambda r$. This suggests widening our perspective to consider all possible radial Hamiltonians with CSI and spherical symmetry. Such Hamiltonians with radial momentum term $\hat{p}^{2N}$ are given by \footnote{In writing \eqref{Eq:GenericHamiltonian0} we have ignored the possibility of $\delta$-function interactions which scale correctly only in certain dimensions. When such terms can be included in \eqref{Eq:GenericHamiltonian0} they can generally be represented as a choice of boundary conditions \cite{Gitman:2009era} rather than being introduced explicitly.}
  \begin{eqnarray}
    \label{Eq:GenericHamiltonian0}
    \hat{H}_{N} = \hat{p}^{2N} + \sum_{i=1}^{2N}  \frac{\lambda_i}{r^i} d_r^{2N-i} 
  \end{eqnarray}
where $N>0$ is an integer, $\lambda_i \in \mathbb{R}$ and we work in units where $m = 1/2$.}
  
{\ Under $r \mapsto \Lambda r$ the Hamiltonians \eqref{Eq:GenericHamiltonian0} scale as $\hat{H}_{N} \mapsto \Lambda^{-2N} \hat{H}_{N}$ making the Schr\"{o}dinger equation scale invariant with $t \mapsto \Lambda^{2N} t$. Anisotropic scaling between space and time is collectively referred to as ``Lifshitz symmetry'' \cite{Alexandre:2011kr}. This scaling symmetry can be seen for example at the finite temperature multicritical points of certain materials \cite{PhysRevLett.35.1678,PhysRevB.23.4615} and in strongly correlated electron systems \cite{PhysRevB.69.224415,PhysRevB.69.224416,Ardonne:2003wa}. Quartic dispersion relations ($E \sim \vec{p}^4$) can also be found in graphene bilayers \cite{0034-4885-76-5-056503} and heavy fermion metals \cite{2012PhRvL.109q6404R} or bose gases \cite{PhysRevA.91.033404,2015PhRvA..91f3634R,2015NatCo...6E8012P,PhysRevB.96.085140}. Lifshitz symmetry may also have applications in particle physics \cite{Alexandre:2011kr}, cosmology \cite{Mukohyama:2010xz} and quantum gravity \cite{PhysRevD.57.971,Kachru:2008yh,Horava:2009if,Horava:2009uw,Gies:2016con}. Moreover, instances of the Hamiltonians in \eqref{Eq:GenericHamiltonian0} can be recovered from Lifshitz field theories coupled to background gauge fields \cite{Das:2009fb,Alexandre:2011kr} of the appropriate multipole moment. Coupling the charged particles to a magnetic monopole in two dimensions, or an infinite solenoid in three, is one way to generate the derivative interactions of \eqref{Eq:GenericHamiltonian0}.}

% What we consider in this paper
{\ In this paper we formulate a RG description for systems described by \eqref{Eq:GenericHamiltonian0}. We show that departure from scale invariance characterised by fixed point annihilation, and subsequently universal DSI, is a generic feature in the landscape of Hamiltonians \eqref{Eq:GenericHamiltonian0}. Depending on the values of $\lambda_i$, we find additional possibilities including: isolated periodic flow (non-linear limit cycle) and quasi-periodic flow (limit tori) as shown by figs.~\ref{Fig:limitcycle} and \ref{Fig:torus} respectively. In addition, we show that these types of RG flows can be simply determined from the characteristic power laws of the $E=0$ wave function (zero modes).}  

\section{CSI in quantum mechanics}

{\ The scaling symmetry of \eqref{Eq:GenericHamiltonian0} implies that if there is one negative energy bound state then there is an unbounded continuum. Thus, the existence of any bound state necessitates that the Hamiltonian is not self-adjoint \cite{Bonneau:1999zq,2015AnHP...16.2367I}. The origin of this phenomenon is the strong singularity of the potential terms at $r=0$. To render the quantum problem well defined we introduce a cut-off $L > 0$ and choose boundary conditions that make the Hamiltonian self-adjoint. The cut-off explicitly breaks scale invariance and we will track the behaviour of the system for $\epsilon L \ll 1$, with $\epsilon = |E|^{1/(2N)}$ and $E$ the energy, using a RG approach.}

{\ An analytic general solution of
  \begin{eqnarray}
    \label{Eq:EigenvalueEqn}
    \hat{H}_{N} \psi(r) = E \psi(r) \; , \qquad r \in [ L, \infty) \;  ,
  \end{eqnarray}
is given in terms of generalized hypergeometric functions \cite{luke1969special} (see supplementary note 1). Importantly, there is an equal number of normalisable eigenfunctions with positive and negative imaginary energies ($N$ to be exact, see supplementary note 1). Therefore, according to von Neumann's second theorem \cite{9780817646622}, there is a $U(N)$ parameter family of self-adjoint boundary conditions at $r=L$ (self-adjoint extensions of $\hat{H}_{N}$). The complete family of boundary conditions is obtained from an impenetrable wall condition corresponding to the vanishing of the radial component of probability current $J(r)$ at $r=L$. In particular,
 \begin{eqnarray}
       J(L) \propto \int_{r=L}^{\infty} dr \; \left[ \psi^{*}(r) \hat{H}_{N} \psi(r) - \psi(r) \hat{H}_{N} \psi^{*}(r)  \right] \; , \; \;
       \label{Eq:probabilitycurrent}
  \end{eqnarray}
where $\hat{H}_{N}$ is a differential operator as given by \eqref{Eq:GenericHamiltonian0} and we absorbed a Jacobian factor into the definition of $\psi(r)$. Using integration by parts on \eqref{Eq:probabilitycurrent} we can reduce this expression to a boundary term. Assuming decay at infinity, this becomes a quadratic form evaluated at $r=L$ in terms of $L^{k-1} d_{r}^{k-1} \psi(L)$ and their conjugates. By diagonalising the quadratic form $J(L)$, it can be reduced to \cite{9780817646622}:
  \begin{eqnarray}
    \label{Eq:ProbabilityCurrentStandardForm}
    J(L) \propto i \left[ | \vec{\psi}^{+}(L) |^2 - | \vec{\psi}^{-}(L) |^2 \right] \; ,
  \end{eqnarray}
where $\vec{\psi}^{\pm}(L)$ are $N$-vectors whose components are linear combinations of the $L^{k-1} d_{r}^{k-1} \psi(L)$. The self-adjoint boundary conditions, being those that set \eqref{Eq:ProbabilityCurrentStandardForm} to zero, are thus
  \begin{eqnarray}
   \label{Eq:BoundaryConditions}
   \vec{\psi}^{+}(L) = U_{N} \vec{\psi}^{-}(L)
  \end{eqnarray}
where $U_{N}$ is an arbitrary $(N \times N)$-matrix \footnote{It is important to note a global symmetry of the boundary conditions. Namely, given some $U_{N}(L)$ defining the boundary conditions then
  \begin{eqnarray}
   \left( V_{1} \vec{\psi}^{+}(L) \right) = \tilde{U}_{N}(L) \left( V_{2} \vec{\psi}^{-}(L) \right) \; , 
  \end{eqnarray}
yields the same boundary conditions where $\tilde{U}_{N}(L) =  V_{1} U_{N}(L) V_{2}^{\dagger}$ for any unitary matrices $V_{1}$, $V_{2}$. The reader therefore, by choosing a different diagonalisation of the probability current $J(L)$, can arrive at a different value for $U_{N}(L)$ to that presented. However, there are still $N^{2}$ distinct $\beta$-functions as this global symmetry only allows us to set the value of $U_{N}(L)$ arbitrarily at a single point along the RG flow.}. The matrix $U_{N}$ describes implicit model dependent parameters that are specified by additional physical information.

\begin{figure}
  \includegraphics[width=0.9\columnwidth]{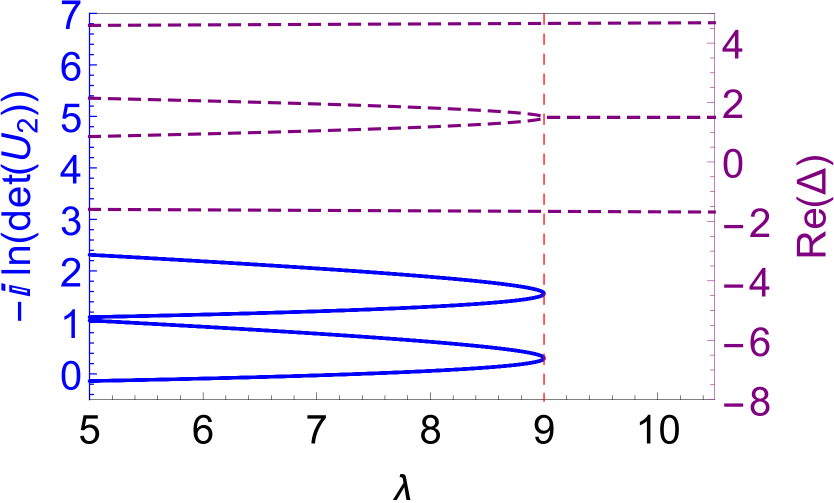}
  \caption{A plot demonstrating fixed point annihilation described by Hamiltonian \eqref{Eq:N=2standardsystem}. The solid blue lines represent the fixed points of the RG flow equation \eqref{Eq:NARE} while the dashed purple lines display the real parts of the roots $\Delta_{i}$; both against the coupling $\lambda$. The RG space is four dimensional and for brevity the fixed points are projected onto a one dimensional axis corresponding to the value of $-i \ln \left( \det U_2 \right) $. The dotted red line indicates the critical coupling $\lambda_{c}=9$ above which there are no powers $\Delta_{i}$ on the line $\Re[z]=N - 1/2=3/2$.}
  \label{Fig:N=2fixedpts}
  \vspace{-1em}
\end{figure} 

{\ As will be exhibited in more detail later, the characteristic low energy behaviour of system \eqref{Eq:EigenvalueEqn} is determined by $2N$ powers $\Delta_i$ describing the $E=0$ eigenfunctions of $\hat{H}_N$. These are obtained by inserting $\psi\propto r^\Delta$ into $\hat{H}_N\psi=0$ and solving for the roots of the resultant polynomial in $\Delta$. Since $\lambda_i \in \mathbb{R}$ in \eqref{Eq:GenericHamiltonian0}, $\Delta_i^\ast$ belongs to the set of roots whenever $\Delta_i$ does. In addition, $\hat{H}_N = \hat{H}^\dagger_N$ implies that $2N-1-\Delta_i$ is also a root (see supplementary note 2). As a result, in the complex $z$ plane, the roots $\Delta_i$ are symmetric with respect to the lines $\Im[z]=0,\,\Re[z]=N-1/2$.}

{\ It will be useful for deriving a RG equation to rewrite \eqref{Eq:EigenvalueEqn} in terms of $\vec{\psi}^{\pm}(r)$ at $r=L$. This consists of splitting \eqref{Eq:EigenvalueEqn} into a set of first order coupled ODEs in $r^{k-1} d_{r}^{k-1} \psi(r)$, $k=1,\ldots, 2N$ and applying the transformation that diagonalised $J$. The result is an equation of the form
  \begin{eqnarray}
    \label{Eq:LinearEOM}
   r d_{r} \left( \begin{array}{c} \vec{\psi}^{+}(r) \\ \vec{\psi}^{-}(r) \end{array} \right)
   &=& \left( \begin{array}{cc}
		C_{++} & C_{+-} \\
		C_{-+} & C_{--} 
              \end{array} \right) \left( \begin{array}{c} \vec{\psi}^{+}(r) \\ \vec{\psi}^{-}(r) \end{array} \right) \; . \qquad
  \end{eqnarray}
Scale invariance ensures that the matrix of $C$s is dimensionless and therefore depends only on $\epsilon r$. The precise form of the $C$s will only be necessary when working with a particular Hamiltonian; to determine qualitative features of the RG space we will not require these details.}

\begin{figure}
  \includegraphics[width=0.9\columnwidth]{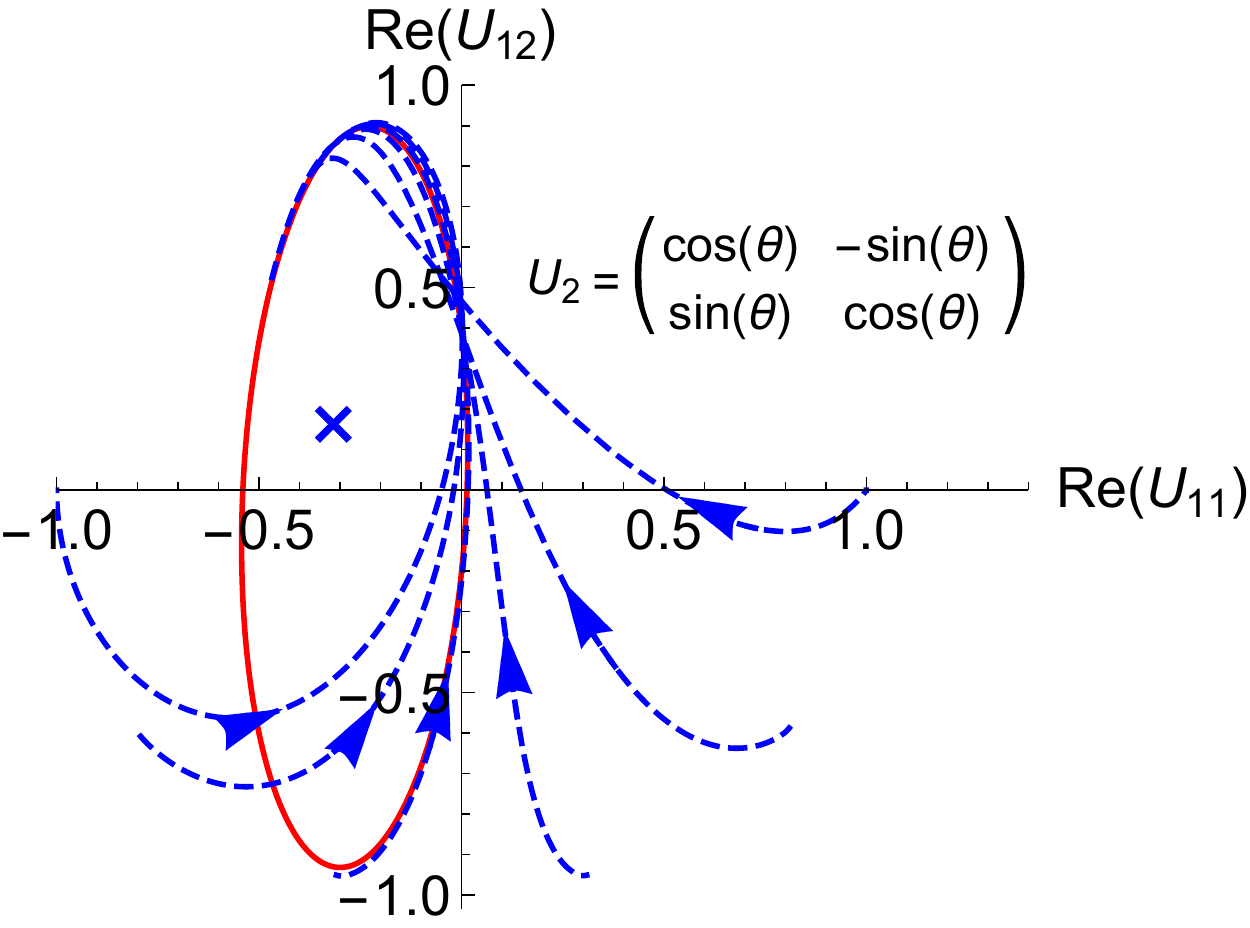}
  \caption{A two dimensional projection of the (four dimensional) RG picture of the system $\hat{H}_{2} = d_{r}^{4}-2/r^4$. Boundary conditions are with respect to $\psi_{1}^{\pm}(L) = \frac{1}{\sqrt{2}} \left( L \psi'(L) \mp i L^2 \psi''(L) \right)$ and $\psi_{2}^{\pm}(L) = \frac{1}{\sqrt{2}} \left( \psi(L) \pm i L^3 \psi'''(L) \right)$. The initial conditions for the dashed blue flows are specified by choosing $\theta=-\pi,\ldots,-\pi/10,0$ for the $U_{2}$ matrix as displayed. We see that all the trajectories flow towards a limit cycle. There exists a non-unitary fixed point, denoted by the blue cross, which is enclosed by the cycle when we project down onto any two dimensional subspace.}
  \label{Fig:limitcycle}
  \vspace{-1em}
\end{figure}

\section{Renormalisation group flow}

{\ Consider an eigenfunction of Hamiltonian \eqref{Eq:GenericHamiltonian0} of energy $E$ and satisfying the boundary condition defined by $U_{N}$ at $r=L$. Defining $U_N \equiv U_N(L)$ and imposing that \eqref{Eq:BoundaryConditions} holds for the given state after performing an infinitesimal transformation $L \mapsto \Lambda L \sim L \left( 1 + dL \right)$ implies that $U_{N}(L)$ must satisfy the following equation:
  \begin{eqnarray}
   L d_{L} \vec{\psi}^{+}(L) = L d_{L} U_{N}(L) \vec{\psi}^{-}(L) + U_{N}(L) L d_{L} \vec{\psi}^{-}(L) \; . \qquad 
  \end{eqnarray}
We replace the derivatives of the field using \eqref{Eq:LinearEOM}, and $\vec{\psi}^{+}(L)$ for $\vec{\psi}^{-}(L)$ using \eqref{Eq:BoundaryConditions}, to find:
  \begin{eqnarray}
    \label{Eq:RGflowintermediate}
   0 &=& \left[ L d_{L} U_{N}(L) - C_{+-} + U_{N}(L) C_{--} - C_{++} U_{N}(L) \right. \nonumber \\
     &\;& \left. \; + U_{N}(L) C_{-+} U_{N}(L) \right]  \vec{\psi}^{-}(L) \; 
  \end{eqnarray}
where $E$ and explicit $L$ dependence enters into \eqref{Eq:RGflowintermediate} through the $C$s. Assuming $\epsilon L \ll 1$ removes this dependence rendering \eqref{Eq:RGflowintermediate} translationally invariant in $E,L$. In this regime, \eqref{Eq:RGflowintermediate} holds for every eigenfunction and its corresponding $\vec{\psi}^{-}(L)$, meaning the term in square brackets is zero. Multiplying through \eqref{Eq:RGflowintermediate} by $- i U_{N}^{-1}$ gives the flow equation for $U_N(L)$:
  \begin{eqnarray}
    \label{Eq:NARE}
   - i L U_{N}^{-1} d_{L} U_{N} &=& i C_{--} - i U_{N}^{-1} C_{+-} + i C_{-+} U_{N} \nonumber \\
     &\;& - i U^{-1}_{N} C_{++} U_{N} \; . 
  \end{eqnarray}
This is essentially a generalisation of the approach taken by \cite{mueller2004renormalization,Kaplan:2009kr}. For $N=1$, defining $g=\tan(- i \ln U_{1})$, equation \eqref{Eq:NARE} reduces into their result.}

\begin{figure}
  \includegraphics[width=0.9\columnwidth]{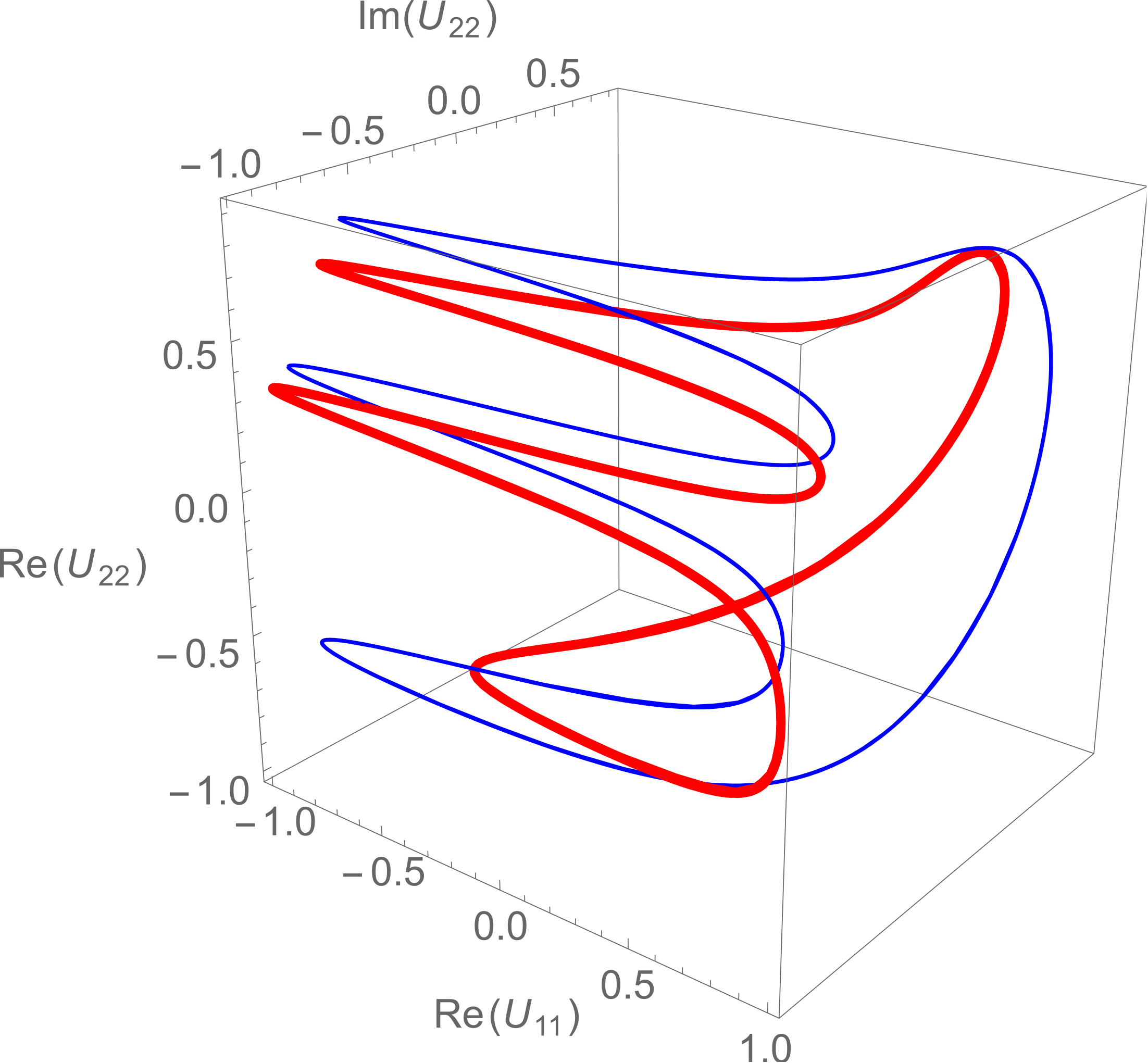}
  \caption{A three dimensional projection of the (four dimensional) RG picture for the system  $\hat{H}_{2} = d_{r}^4 +\frac{25}{2} \frac{1}{r^2} d_{r}^2 - 25 \frac{1}{r^3} d_{r} +\frac{585}{16} \frac{1}{r^4}$. The boundary conditions are $U_{2}=-\mathbbm{1}_{2}$ (red and thick) and $U_{2}=\exp (-3 \pi i/4) \mathbbm{1}_{2}$ (blue and thin) at $L=e^{1}$ with respect to the basis $\psi_{1}^{\pm}(L) \approx \mp 0.016 i \psi(L) - 0.016 L \psi'(L) \pm 0.199 i L^2 \psi''(L)  + 0.199 L^3 \psi'''(L)$ and $\psi_{2}^{\pm}(L) \approx \mp 2.500 i \psi(L) + 2.500 L \psi'(L) \mp 0.199 i L^2 \psi''(L)  + 0.199 L^3 \psi'''(L)$. We see that a small modification of the boundary condition corresponds to two different nearby periodic trajectories. The space is filled by closed trajectories. Any choice of initial condition, that is the initial boundary condition, will flow on one of them.}
  \label{Fig:limitcycle2}
  \vspace{-1em}
\end{figure}

\section{Fixed point annihilation -- a generic feature in the landscape of scale invariant Hamiltonians }

{\ We numerically obtained the trajectories corresponding to \eqref{Eq:NARE} and solved for the zeros of the RHS (the $\beta$ function) in a variety of cases. We find a range of distinct flows terminating in fixed points, limit cycles and limit tori. We find that a few simple properties of the $E=0$ power laws $\Delta_i$ determine what is the characteristic RG picture as summarised in table \ref{tab:features}. In particular, the RG space will contain unitary fixed points if and only if there are no roots $\Delta_i$ on the symmetry line $\Re[z] = N-1/2$. This implies the following general result: consider a Hamiltonian $\hat{H}_N$ corresponding to some choice of $\lambda_i \in \mathbb{R}$ in \eqref{Eq:GenericHamiltonian0} such that there are no roots $\Delta_i$ on the symmetry line $\Re[z] = N-1/2$. Then, continuously tuning the $\lambda_i$'s such that at least one pair of roots settle on $\Re[z] = N-1/2$ will generate a transition characterized by fixed point annihilation. In this context, fixed point annihilation of  $\hat{H}_S= p^2/2m - \lambda_2/r^2$ is only one case, corresponding to $N = 1$, $\lambda_1 = 0$. In general we observe that there are $2^N$ unitary fixed points which annihilate in pairs.}

\begin{figure}
  \includegraphics[width=0.85\columnwidth]{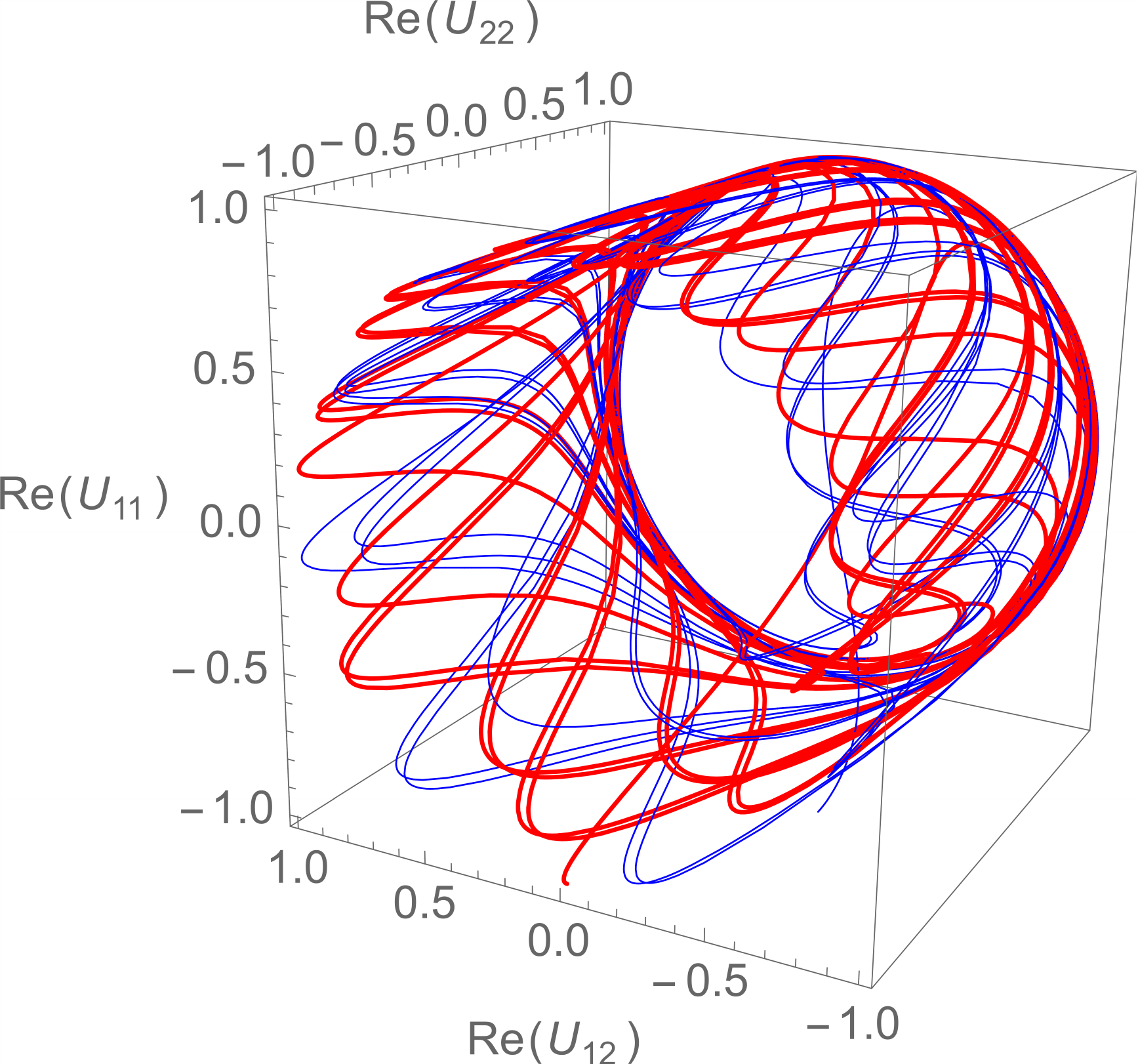}
\caption{A three dimensional projection of the (nine dimensional) RG picture for the system  $\hat{H}_{3} \approx - d_{r}^6 - 16.2 \frac{1}{r^2} d_{r}^4 + 64.9 \frac{1}{r^3} d_{r}^3 - 195.3 \frac{1}{r^4} d_{r}^2 + 392.0 \frac{1}{r^5} d_{r} - 299.1 \frac{1}{r^6}$. The basis with respect to which we determine $U_{3}$ is particularly long, and as such we display it in supplementary note 4 along with an exact expression for $\hat{H}_{3}$. The initial condition for the red, thick curve is $U_{3}(1)=-\mathbbm{1}_{3}$ and $U_{3}(1)=\exp(i \pi/4) \mathbbm{1}_{3}$ for the blue and thin curve. The curves represent two different initial conditions that are attracted to a quasi periodic trajectory as $L \rightarrow 0$. This type of attractor is characterised by a trajectory that never closes on itself and fills a compact RG subspace.}
  \label{Fig:torus}
  \vspace{-1em}
\end{figure}

{\ As an example, consider fig.~\ref{Fig:N=2fixedpts} which represents the flow of the fixed points of \eqref{Eq:NARE} for the system:
  \begin{eqnarray}
    \label{Eq:N=2standardsystem}
    \left[ \left( d_{r}^2 + \frac{1}{r} d_{r} - \frac{m^2}{r^2} \right)^2 - \frac{\lambda}{r^4} - E \right] \left( r^{-\frac{1}{2}} \psi(r) \right) = 0 \;  \qquad
  \end{eqnarray}
which describes a particle with kinetic energy $E = \vec{p}^4$ on a two-dimensional plane interacting with a potential whose strength is controlled by the parameter $\lambda$. The integer $m$ represents the angular momentum while the additional factor of $r^{-1/2}$ is a Jacobian factor such that the probability current is defined as in \eqref{Eq:probabilitycurrent}. Choosing $m=2$ henceforth, the boundary conditions are specified by $U_{2}$ matrices with respect to the basis
  \begin{subequations}
    \begin{eqnarray}
      \label{Eq:N=2standardbcs1}
      \psi^{\pm}_{1}(L) &\approx& 0.033 \, \psi(L) \pm 0.033 i \, L \psi'(L) \nonumber \\
			&\;& + 0.254 \, L^2 \psi''(L) \pm 0.254 i \, L^3 \psi'''(L)  \; , \qquad \\
      \label{Eq:N=2standardbcs2}
      \psi^{\pm}_{2}(L) &\approx& 1.937 \, \psi(L) \mp 1.937 i \, L \psi'(L) \nonumber \\
			&\;& - 0.254 \, L^2  \psi''(L) \pm 0.254 i \, L^3 \psi'''(L) \; . \qquad
    \end{eqnarray}
  \end{subequations}
Different values of $m$ will yield different numerical coefficients in \eqref{Eq:N=2standardbcs1} and \eqref{Eq:N=2standardbcs2}, as each choice of $m$ in \eqref{Eq:N=2standardsystem} corresponds to a distinct Hamiltonian.}

{\ For $\lambda < 9$, there are four unitary fixed points (and a further two non-unitary). When $\lambda>\lambda_{c} = 9$, the red dotted line of fig.~\ref{Fig:N=2fixedpts}, there are no unitary fixed points. In terms of the roots $\Delta_i$, the value $\lambda_{c} = 9$ is the exact point at which roots move onto the symmetry line $\Re[z]=3/2$, as seen in fig.~\ref{Fig:N=2fixedpts}. An additional illustration of this phenomenon for $N=3$ is given in supplementary note 3.}

\begin{figure}
  \includegraphics[width=0.95\columnwidth]{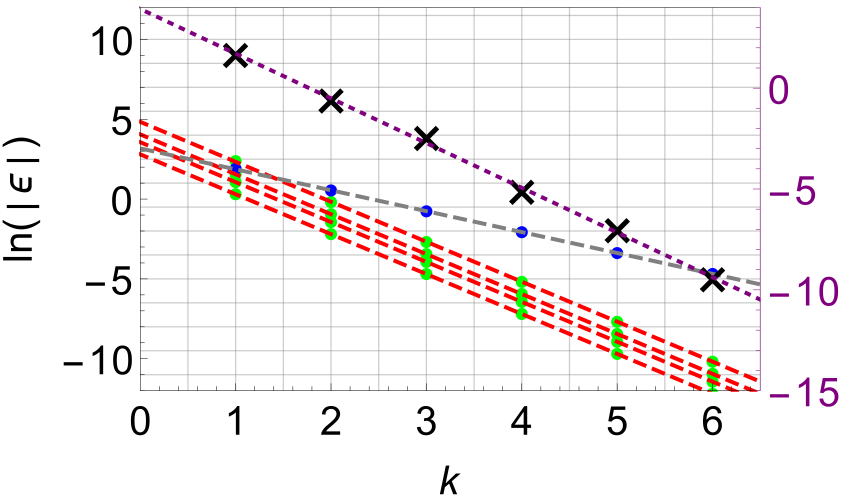}
  \caption{Negative bound state energies of various Hamiltonians corresponding to the distinct flows described by figs.~\ref{Fig:limitcycle}, \ref{Fig:limitcycle2}, \ref{Fig:torus}. The blue and green dots represent bound state energies for the Hamiltonians $\hat{H}_{2}= d_{r}^4 + \frac{\left(125+388 \pi ^2\right)}{50} \frac{1}{r^2} d_{r}^2 - \frac{\left(125+388 \pi ^2\right)}{25} \frac{1}{r^3} d_{r} + \frac{\left(9+4 \pi ^2\right) \left(225+676 \pi ^2\right)}{400} \frac{1}{r^4}$ and \eqref{Eq:N=2standardsystem} (with $\lambda=100$ in this latter case). The black crosses represent negative energy levels for the Hamiltonian of fig.~\ref{Fig:torus}. All boundary condition parameters are given by the identity matrix and the cut-off is $L=e^{-1}$. The first pair of these systems are of the class corresponding to figs.~\ref{Fig:limitcycle} and \ref{Fig:limitcycle2} respectively. In the former, the spectrum is a composition of four intertwined geometric towers of energy (the four red dashed lines), while in the latter there is one (the grey dashed line). The Hamiltonian corresponding to the black crosses is a limit torus case and The purple dotted line represents a best fit to the data, given by black crosses. These crosses do not sit precisely on the purple line indicating that there is no discrete scale invariance.}
  \label{Fig:energytowers}
  \vspace{-1em}
\end{figure}

{\ When considering the phenomena of fixed point annihilation, a pertinent question is what is the characteristic RG picture in the over critical regime, i.e.~the regime with no fixed points. Recent studies \cite{BraatenPhysRevA.70.052111,HammerSwingle2006a, GorskyPhysRevD.89.061702, NishidaPhysRevB.94.085430} show that for $N=1$ the  flow in the over critical regime is completely periodic. In other words, regardless of the initial condition, the boundary parameter is periodic in $\log L$ generating a DSI RG picture. The appearance of this type of flow has been considered as evidence for the relevance of RG limit cycles in physical applications.}

{\ For $N>1$, we find that the $N=1$ case is a single instance in a rich set of possibilities. In the overcritical regime, and close to the critical point, there is an isolated closed trajectory to which all other trajectories are attracted as $L \rightarrow 0$ (see fig.~\ref{Fig:limitcycle}). As opposed to completely periodic flow, this intrinsically non-linear flow picture, is in fact the rigorous definition of a limit cycle \cite{StrogatzNonlinear}. To our knowledge, this is the only manifestation of a limit cycle in a physical application to date. The difference with respect to the $N=1$ case is simply displayed in terms of the behaviour of the $E=0$ wave functions, i.e., the roots $\Delta_i$. For $N>1$, near the critical point and in the overcritical regime, the two complex conjugate roots on the symmetry line $\Re[z] = N-1/2$ are accompanied by $2(N-1) \neq 0$ roots off the line. If we move in a direction in the $\lambda_i$ parameter space such that all the roots are on the symmetry line, the limit cycle will disappear in favour of an RG space filled entirely by periodic flows (fig.~\ref{Fig:limitcycle2}) or quasi-periodic flows. The former is obtained when the imaginary part of all the roots on the symmetry line has a common divisor and later when they don't. If we allow roots outside the symmetry line as well as multiple roots on the symmetry line (with imaginary parts not having a common divisor), then all the flows are attracted to an isolated quasi-periodic trajectory as seen in fig.~\ref{Fig:torus}. This trajectory, known as a limit torus, is characterised by a curve that never closes on itself and fills a compact RG  subspace.}

{\ In order to obtain further insight on the over critical regime, we calculated the spectrum in various cases corresponding to the distinct flows described by figs.~\ref{Fig:limitcycle}, \ref{Fig:limitcycle2}, \ref{Fig:torus}. For $\hat{H}_S= p^2/2m - \lambda/r^2$, corresponding to $N = 1$, $\lambda_1 = 0$ and $\lambda_{2}=-\lambda$, DSI manifests in the geometric progression of the spectrum given by \eqref{Eq:geometric spectrum}. For $N>1$ and in the case where the flow is periodic (figs.~\ref{Fig:limitcycle}, \ref{Fig:limitcycle2}) we find that the spectrum can be described by a union of multiple geometric towers as seen for example in fig.~\ref{Fig:energytowers}. When the flow is quasi-periodic the spectrum is no longer DSI as is also exhibited in fig.~\ref{Fig:energytowers}. }

{\ We considered a large class of quantum mechanical scale invariant systems \eqref{Eq:GenericHamiltonian0} and formulated a RG description controlled by a short distance cut-off $L$. The resulting picture shows that the quantum phase transition characterised by fixed point annihilation and DSI is a generic phenomenon exhibited by the class of Hamiltonians \eqref{Eq:GenericHamiltonian0}. We found that the transition point is related to the value of the roots characterising the zero energy wave function solutions. Hermiticity of the Hamiltonian imposes that these powers be symmetric with respect to the line $\Re[z] = N-1/2$ and the appearance of roots on this line is in direct correspondence with the transition point. We hope that our results will provide further insight and intuition on the quantum behaviour of scale invariant systems in quantum mechanics and quantum field theory.}

\begin{acknowledgments}
{\ The work of DB is supported by key grants from the NSF of China with grant numbers:~11235010 and 11775212. This work was also supported by the Israel Science Foundation Grant No.~924/09. DB would like to thank Matteo Baggioli for reading an early draft.}
\end{acknowledgments}

%%%%%%%%%%%% Bibliography %%%%%%%%%%%%

\bibliography{references}

% %%%%%%%%%%%% Supplementary Material %%%%%%%%%%%%
% % 
\clearpage


\section*{Supplementary material}

# On the landscape of scale invariance in quantum mechanics: supplementary material

Daniel K. Brattan[*]

*Interdisciplinary Center for Theoretical Study, University of Science and Technology of China, 96 Jinzhai Road, Hefei, Anhui, 230026 PRC.*

Omrie Ovdat[†] and Eric Akkermans[‡]

*Department of Physics, Technion Israel Institute of Technology, Haifa 3200003, Israel.*

(Dated: May 15, 2018)

## Supplementary Note 1: Analytic solution of $\hat{H}_N \Psi(r) = E \Psi(r)$

Consider the differential equation $\hat{H}_N \Psi(r) = E \Psi(r)$ where

$$\hat{H}_N = \hat{p}^{2N} + \sum_{i=1}^{2N} \frac{\lambda_i}{r^i} d_r^{2N-i}, \qquad (S1)$$

where $\lambda_i \in \mathbb{R}$. For $E = 0$, there are $2N$ independent solutions given by $\psi \propto r^{\Delta_i}$, $i = 1, \ldots, 2N$ obtained by inserting $\psi \propto r^\Delta$ into $\hat{H}_N \psi = 0$ and solving for the roots of the resultant polynomial in $\Delta$

$$0 = (\Delta - \Delta_1) \ldots (\Delta - \Delta_{2N}) . \qquad (S2)$$

We assume that all roots are distinct for simplicity as this avoids dealing with logarithms in the Frobenius solution. For $E \neq 0$ and of arbitrary complex value, the general solution of (S1) is expressed in terms of generalized hypergeometric functions ${}_0F_{2N-1}$ [S1]:

$$\psi(r; \phi_j) = \sum_{i=1}^{2N} e^{i\Delta_i \phi_j} \left(\frac{\epsilon r}{2N}\right)^{\Delta_i} \Gamma\left(\frac{\boldsymbol{\Delta}_i - \Delta_i}{2N}\right)$$

$$ {}_0F_{2N-1}\left[\begin{array}{c} - \\ 1 - \frac{\boldsymbol{\Delta}_i - \Delta_i}{2N} \end{array}; |E|\left(\frac{r}{2N}\right)^{2N}\right], \quad (S3)$$

$$\phi_j = \frac{1}{2N}(\theta - 2\pi(N + (w + j))) ,$$
$$j = 1, 2, \ldots, 2N , \qquad (S4)$$

where $\boldsymbol{\Delta}_i$ is the vector of solutions to (S2) with $\Delta_i$ omitted, $w$ is an integer chosen such that $|\phi_j| < \pi(1 + \frac{1}{2N})$, $E = |E|e^{i\theta}$ and $\epsilon = |E|^{\frac{1}{2N}}$.

At large $r$ only the leading derivative term $(-1)^N d_r^{2N}$ of the kinetic term is important and the wavefunctions behave as $\psi(r) \sim \exp(\alpha r)$ with $\alpha^{2N} = (-1)^N E$. Setting $E = \pm i$ we find

$$\alpha = \exp\left(i\pi\left(\frac{1}{2} + \frac{n}{N} \pm \frac{1}{4N}\right)\right) , \qquad (S5)$$

respectively with $n = 0, \ldots, 2N - 1$. For each $E = \pm i$ there are exactly $N$ values of $n$ giving $\pi/2 < \arg(\alpha) < 3\pi/2$. Thus one has $N$ decaying wavefunctions of negative and positive imaginary energies giving a $U(N)$

self-adjoint extension. The relationship between these asymptotic behaviours and (S3) is:

$$\psi(r; \phi_j) \sim \exp\left(-\epsilon r e^{i\phi_j}\right) \qquad (S6)$$

for non-zero $\epsilon$ (see [S1]).

## Supplementary Note 2: Constraints on the powers laws characterising the $E = 0$ wavefunctions

The space of wave functions upon which $\hat{H}_N$ acts is equipped with an inner product defined given by

$$\langle \phi | \psi \rangle = \int_{r=L}^{\infty} dr \, \phi^*(r) \psi(r) \qquad (S7)$$

where we have absorbed a Jacobian factor into the definition of the wave functions and $L \geq 0$. Hamiltonian (S1) can be factorised as

$$\hat{H}_N = (-1)^N \hat{D}_{2N} \ldots \hat{D}_1 \qquad (S8)$$

where:

$$\hat{D}_i = d_r - \left(\frac{\Delta_i - i + 1}{r}\right) , \qquad (S9)$$

$i = 1, \ldots, 2N$, and $\Delta_i$ are the roots of (S2). This can be seen by acting with the $\hat{D}_i$'s on $r^\Delta$, $\Delta$ arbitrary. The formal adjoints of (S9), are defined by $\langle \phi | \hat{D}_i^\dagger | \psi \rangle = \langle \hat{D}_i \phi | \psi \rangle$, ignoring any boundary terms. By integrating by parts the right hand side, $\hat{D}_i^\dagger$ is given by

$$\hat{D}_i^\dagger = -\left[d_r + \left(\frac{\Delta_i^* - i + 1}{r}\right)\right]$$
$$= -\left[d_r - \left(\frac{(2N - 1 - \Delta_i^*) - (2N - i + 1) + 1}{r}\right)\right] .$$

Similarly, the formal adjoint of $\hat{H}_N$ is given by

$$\hat{H}_N^\dagger = (-1)^N \hat{D}_1^\dagger \ldots \hat{D}_{2N}^\dagger . \qquad (S10)$$

For $\hat{H}_N = \hat{H}_N^\dagger$ we can identify $\hat{D}_i^\dagger = -\hat{D}_{2N-i+1}$. As a result, if $\Delta_i$ belongs to the set $\{\Delta_i\}$ then so does $2N - 1 - \Delta_i^*$. Furthermore, since $\lambda_i$ in (S1) are real, both $\Delta_i$ and $\Delta_i^*$ belong to the set of roots. Thus, if $\Delta_i^*$ is a root, then $2N - 1 - \Delta_i$ is a root which implies that if $\Delta_i$ is a root then $2N - 1 - \Delta_i$ is a root.

**Supplementary note 3: Fixed point flow for $N = 3$**

Consider fig. S1 which represents the flow of the fixed points of (S15) for the system:

$$\left[\left(d_r^2 + \frac{1}{r}d_r - \frac{m^2}{r^2}\right)^3 + \frac{\lambda}{r^6} + E\right]\left(r^{-\frac{1}{2}}\psi(r)\right) = 0 \,. \quad \text{(S11)}$$

This represents a particle with kinetic energy $E = \boldsymbol{p}^6$ on a two-dimensional plane interacting with a potential whose strength is controlled by the parameter $\lambda$. The integer $m$ represents the angular momentum while the additional factor of $r^{-1/2}$ is a Jacobian factor such that the probability current is defined as in equation (4) of the main text. Choosing $m = 1$ henceforth, the boundary conditions are specified by $U_3$ matrices with respect to the basis

$$\psi_1^\pm(L) \approx \mp 0.0067i\,\psi(L) + (0.0066 \mp 0.0001i)\,L\psi'(L) - (0.0014 \pm 0.0159i)\,L^2\psi''(L) + (0.0156 \mp 0.0307i)\,L^3\psi'''(L)$$
$$- (0.0031 \pm 0.1461i)\,L^4\psi^{(4)}(L) + 0.1493\,L^5\psi^{(5)}(L)\,, \quad \text{(S12a)}$$
$$\psi_2^\pm(L) \approx (0.0049 \pm 0.1020i)\,\psi(L) - (0.1004 \pm 0.2103i)\,L\psi'(L) + (0.0111 \mp 0.7618i)\,L^2\psi''(L) + 0.7731\,L^3\psi'''(L)$$
$$- (0.1640 \mp 0.0783i)\,L^4\psi^{(4)}(L) - (0.0795 \mp 0.0039i)\,L^5\psi^{(5)}(L)\,, \quad \text{(S12b)}$$
$$\psi_3^\pm(L) \approx 2.9105\,\psi(L) + (0.1013 \mp 2.8055i)\,L\psi'(L) + (0.3449 \pm 0.7747i)\,L^2\psi''(L) - (0.0449 \pm 0.3595i)\,L^3\psi'''(L)$$
$$- (0.1626 \pm 0.0059i)\,L^4\psi^{(4)}(L) \pm 0.1687i\,L^5\psi^{(5)}(L)\,. \quad \text{(S12c)}$$

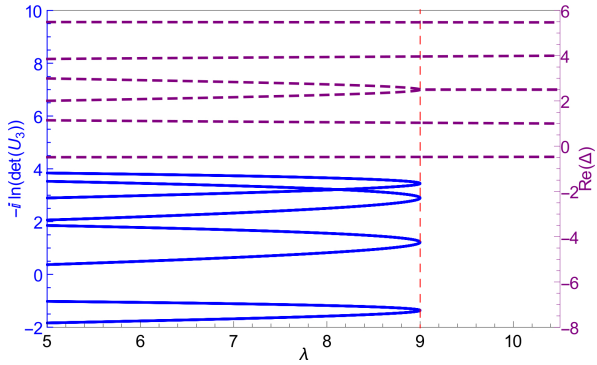

FIG. S1. A plot demonstrating fixed point annihilation for a two dimensional system with orbital angular momentum ($m = 1$) interacting with a dipole potential (S11). The solid blue lines represent the fixed points of the RG flow equation (S15) while the dashed purple lines display the real parts of the roots $\Delta_i$; both against the dipole coupling $\lambda$. The dotted red line indicates the critical coupling $\lambda_c = 0$ above which there is DSI.

Different values of $m$ will yield different numerical coefficients in (S12a), (S12b) and (S12c), as each choice of $m$ in (S11) corresponds to a distinct Hamiltonian. For $\lambda < 9$, there are 8 unitary fixed points. When $\lambda > \lambda_c = 9$, the red dotted line of fig. S1, there are no unitary fixed points. In terms of the roots $\Delta_i$, the value $\lambda_c = 9$ is the exact point at which roots move onto the symmetry line $\text{Re}[z] = 5/2$, as seen in fig. S1.

**Supplementary note 4: Hamiltonians and boundary conditions for $N = 3$ limit tori**

The Hamiltonian that represents a limit torus flow in fig.4 of the main text is:

$$\hat{H}_3 = -d_r^6 - \left(\frac{\pi^2}{4} + \frac{55}{4}\right)\frac{1}{r^2}d_r^4 + \left(\pi^2 + 55\right)\frac{1}{r^3}d_r^3$$
$$- \left(\frac{39\pi^2}{8} + \frac{2355}{16}\right)\frac{1}{r^4}d_r^2 + \left(\frac{27\pi^2}{2} + \frac{1035}{4}\right)\frac{1}{r^5}d_r$$
$$- \left(\frac{549\pi^2}{64} + \frac{13725}{64}\right)\frac{1}{r^6}\,. \quad \text{(S13)}$$

The basis for the $U_3$ matrices describing the boundary conditions are given by:

$$\psi_1^\pm(L) \approx \mp 0.000318i\,\psi(L) + (0.000312 \mp 0.000006i)\,L\psi'(L) + (0.000061 \pm 0.005192i)\,L^2\psi''(L)$$
$$- (0.005096 \mp 0.010420i)\,L^3\psi'''(L) - (0.000996 \pm 0.053012i)\,L^4\psi^{(4)}(L) + 0.054026\,L^5\psi^{(5)}(L), \quad \text{(S14a)}$$
$$\psi_2^\pm(L) \approx -(0.0016482 \pm 0.137145i)\,\psi(L) - (0.132819 \mp 0.276935i)\,L\psi'(L) + (0.026730 \pm 0.836773I)\,L^2\psi''(L)$$
$$+ 0.864357\,L^3\psi'''(L) + (0.172537 \pm 0.082749i)\,L^4\psi^{(4)}(L) + (0.085444 \mp 0.0010269i)\,\psi^{(5)}(L), \quad \text{(S14b)}$$
$$\psi_3^\pm(L) \approx 7.185486\,\psi(L) + (0.346217 \pm 6.821721i)\,L\psi'(L) + (1.056720 \mp 2.257285i)\,L^2\psi''(L)$$
$$+ (0.021347 \pm 1.111986i)\,L^3\psi'''(L) + (0.064511 \mp 0.003274i)\,L^4\psi^{(4)}(L) \pm 0.067951i\,\psi^{(5)}(L). \quad \text{(S14c)}$$

## Fixed point annihilation - methods and additional results

We have used two methods to check that our numerical calculations give the correct values for the stable fixed point. The first of these requires setting the left hand side of the RG-flow equation,

$$-iLU_N^{-1}d_L U_N = iC_{--} - iU_N^{-1}C_{+-} + iC_{-+}U_N \\ -iU_N^{-1}C_{++}U_N \ , \quad (S15)$$

to zero and solving the resultant matrix polynomial for $U_N$. For $N = 1$ this is simply solving a quadratic equation. For $N > 1$ we have a matrix quadratic equation which is more difficult to solve. Nonetheless the methods discussed in [S2] allow one to do this and we refer the interested reader to the book.

A second, more practical, method is to use the general form of the wavefunction for $\epsilon L \ll 1$. The generic wavefunction for small $\epsilon L \ll 1$ has the form

$$\psi(r) = \sum_{i=1}^{N} \left[ (\epsilon r)^{\Delta_i} \phi_i + \ldots + (\epsilon r)^{\Delta_{i+N}} O_i + \ldots \right] \ , \quad (S16)$$

where $\phi_i$ and $O_i$ are complex constants and $\epsilon = |E|^{1/N}$. Half of these coefficients will be fixed by conditions at infinity while the other half will be fixed by boundary conditions on the cut-off.

As the wavefunction $\psi(r)$ can be expanded in small $\epsilon r$ so can $\boldsymbol{\psi}^\pm(r)$. Hence the boundary condition at $r = L$ takes the form

$$0 = \sum_{i=1}^{N} (\epsilon L)^{\Delta_i} \phi_i \left( \boldsymbol{\psi}^+_{\Delta_i} - U_N \boldsymbol{\psi}^-_{\Delta_i} \right) + \ldots \quad (S17) \\ + (\epsilon L)^{\Delta_i + N} O_i \left( \boldsymbol{\psi}^+_{\Delta_{i+N}} - U_N \boldsymbol{\psi}^-_{\Delta_{i+N}} \right) + \ldots$$

where $\epsilon L \ll 1$ and $\boldsymbol{\psi}^\pm_{\Delta_i}$ is the coefficient of $\phi_i(\epsilon L)^{\Delta_i}$ or $O_i(\epsilon L)^{\Delta_{i+N}}$ in the expansion of $\boldsymbol{\psi}^\pm(L)$.

Solutions to the energy eigenvalue problem at some $r = L$ satisfy (S17). Suppose we require (S17) to be satisfied for every $L$ such that $\epsilon L \ll 1$. Then the terms in (S17) must vanish separately. Moreover, after fixing conditions at $r = L$ we will require $N$ remaining degrees of freedom (a collection of $N$ of the $\phi_i$ and $O_i$) to fix boundary conditions at infinity. Thus, it must be the case that $N$ of the $\phi_i$ and $O_i$ are identically zero while for the remaining $N$ terms the equation $\left( \boldsymbol{\psi}^+_{\Delta_i} - U_N \boldsymbol{\psi}^-_{\Delta_i} \right) \equiv 0$ is satisfied (so that $N$ of the $\phi_i$ and $O_i$ are undetermined). These latter conditions allow for a total of $2N!/N!^2$ solutions of $U_N$, each of which is a fixed point (as it is invariant under rescaling).

We have numerically checked the number of fixed points using the method of [S2]. It should be noted that while (S17) seemingly yields $(2N)!/N!^2$ possible fixed points, not all of them are unitary. In particular, we have seen that if one chooses to include two $\Delta_i$ that sum to $2N - 1$ in the definition of $U_N$ then the resultant fixed point will be non-unitary. Thus, we find that there always $2^N$ unitary fixed points. We tested one thousand uniformly distributed random values for $\Delta_i$ with $\mathrm{Re}[\Delta_i], \mathrm{Im}[\Delta_i] \in [-10, 10]$ (satisfying the constraints explained in supplementary note 2) and determined the number of unitary fixed points for $N = 2, 3$ using the method of [S2]. Indeed, applying this method in the case where there are roots on the line $\mathrm{Re}[z] = N - 1/2$ yields only non-unitary fixed points and $2^N$ unitary fixed points when there is no root with $\mathrm{Re}[z] = N - 1/2$.

---

* danny.brattan@gmail.com
† somrie@campus.technion.ac.il
‡ eric@physics.technion.ac.il



\end{document}